\documentclass[a4paper,12pt]{article} 
\usepackage{amssymb}
\usepackage{amsfonts}
\usepackage{graphicx}
\usepackage{psfrag}
\usepackage{epsfig}
\usepackage{latexsym}

\newcommand{\beq}{\begin{equation}}
\newcommand{\eeq}{\end{equation}}
\newcommand{\bea}{\begin{eqnarray}}
\newcommand{\eea}[1]{\label{#1}\end{eqnarray}}                                
\newcommand{\bdm}{\begin{displaymath}}
\newcommand{\edm}{\end{displaymath}}
\newcommand{\beqa}{\begin{eqnarray}}
\newcommand{\eeqa}{\end{eqnarray}}
\newcommand{\beqab}{\begin{eqnarray*}}
\newcommand{\eeqab}{\end{eqnarray*}}
\newcommand{\de}{\partial}
\newcommand{\rf}{\rfloor}
\newcommand{\lf}{\lfloor}
\newcommand{\h}{\frac{1}{2}}
\newcommand{\ar}{\arrowvert_0}

 \def\@makefnmark{\hbox to 0pt{$^{\@thefnmark}$\hss}}  

\newcounter{saveeqn}%

\begin{document}


 \vspace{0.5cm}
 \begin{center}
 {\LARGE \bf  Strong Coupling vs. 4-D Locality in Induced Gravity}\\
 \vspace{0.7cm} 

 \vspace{0.5cm}
 {\large  
  M. Porrati \footnote{e-mail:mp9@SCIRES.ACF.nyu.edu} and 
J.-W. Rombouts \footnote{e-mail: jwr218@nyu.edu}} \\
 \vspace{0.5cm}
 {\it Department of Physics, New York University,\\
 4 Washington Place, New York, NY 10003, USA} \\
\date{\today}
\vspace{1.5cm}
 \end{center}

 \bigskip
\begin{center}
{\bf \large Abstract}
\end{center}
\begin{quotation}\noindent
We re-examine the problem of strong coupling in a regularized version of DGP 
(or ``brane-induced'') gravity. We find that the regularization of ref. 
hep-th/0304148 differs from DGP in that it 
does not exhibit strong coupling or ghosts up to cubic order 
in the interactions. We suggest that the nonlocal nature of the theory, when 
written in terms of the 4-D metric, is 
a plausible reason for this phenomenon. Finally, we
briefly discuss the possible behavior of the model at higher-order in 
perturbation theory.    
\end{quotation}


 \newpage

\section{Introduction}
Modifications of gravity in the UV or IR provide an interesting arena for
 addressing phenomenological issues
ranging from the hierarchy problem to the cosmological constant problem. 
In recent years, several ideas were
proposed for a consistent modification of gravity in the infrared (\cite{first}-\cite{dgs}). 
In particular, the DGP \cite{dgp} model is a five-dimensional brane-world realization  
of a theory which gives modifications of four-dimensional gravity in the far IR. 
 
Higher codimensional generalizations of DGP were considered in \cite{Dvali:2000xg}-\cite{Dubovsky}; 
they differ from the
codimension-one model in several important aspects. For instance, they
 need UV regularization in order to properly define the propagation of 
gravity, since their Green's functions are UV divergent. 
Thus, it seems useful to study generalizations of these models
which still share the same modifications of gravity in the IR, but are regularized 
in the UV.  
In \cite{KPR}, we proposed a regularization procedure for the 
higher-codimension theories, 
which produces a ghost- and tachyon-free spectrum at linear order. 
The basic ingredient is to use
purely higher dimensional gravity, with a region around the brane where the gravitational
coupling is smaller (this may be dynamically realized by coupling gravity to a soliton;
for discussions see refs.\cite{dgp}, \cite{KPR}-\cite{dvali}). 
One can consider both smooth and sharp realizations of this mechanism; in this
 paper we shall confine ourselves to the sharp version, 
for which the action takes the form:
\beq
\label{sd}
 {S/M_*^{2+N}}= 
\epsilon \int_{ \rho< \Delta } d^{4+N} x \sqrt{G_{4+N}}   
{\cal R}_{4+N}+ \int_{ \rho> \Delta}
 d^{4+N} x  \sqrt{G_{4+N}} {\cal R}_{4+N},
\eeq
with $\epsilon$ a large dimensionless quantity and $M_*$ the fundamental Planck scale. 
One can add to this a brane term at the origin of transverse space,
on which the Standard Model (SM) particles are confined. 
When one looks at the gravitational interaction between SM sources 
confined to the 
brane, using the Kaluza-Klein decomposition of the graviton, one sees that this action 
gives rise to four-dimensional behavior, 
within a critical distance scale and higher dimensional behavior above that 
scale \cite{KPR}. For N=1 this behavior coincides with DGP.

In this paper, we further investigate this model in a somewhat different context.
As was discussed in \cite{a9} (see also \cite{Rubakov}), the DGP model
suffers from a problem also encountered in massive gravity, namely that the UV 
cutoff beyond which the effective field theory description breaks down is 
much lower than one would naively expect. 
In way reminiscent to how longitudinal polarizations of a 
spin one field with mass $m_A$ become
strongly coupled at $\Lambda_1= 4 \pi m_A/g$, the longitudinal polarizations of a 
graviton with given 
mass $m_g$ become strongly coupled at a scale $\Lambda_5=(m_g^4 M_P)^{1/5}$. 

A very useful method to find this scale in massive gravity using the 
Goldstone formalism was introduced in \cite{a0}. 
In analogy with gauge theory, the authors identify the 
Goldstone bosons of broken coordinate invariance, 
which encode in a very simple way the scale at which 
the effective field theory description breaks
down. These Goldstone modes are characterized in the Lagrangian formulation 
by fields with anomalously 
small kinetic terms (which, unlike the spin one case, 
vanish in the decoupling limit) and large self-interactions. 
For massive gravity, the Goldstone modes coincide with the 
longitudinal modes of the graviton at high energies. This is nothing but
the gravitational version of the Goldstone equivalence theorem.
In DGP, the Goldstone mode was identified to be the brane-bending mode which 
becomes strongly interacting at a scale $\Lambda_{DGP}= (M_P/L^2)^{1/3}$ \cite{a9},
 where 
$L$ is the DGP scale; $L=M_P^2/M_*^3$ . 
For typical DGP parameters ($L$ of order of the Hubble scale), $\Lambda_{DGP}^{-1}$ is
around $1000$ km, which is indeed unacceptable from the experimental 
point of view. 
At this scale, from the effective field theory point of view, one loses control of the
theory because one has to include all operators consistent with the a priori symmetries  
of the model. This feature is therefore inherently a 
problem of quantum stability.  

At the classical level on the other hand, this problem translates into 
the appearance of the
"Vainshtein scale" \cite{vain}. Vainshtein noted that in massive gravity, the linear 
approximation near a heavy source breaks down at a distance
\beq
d_v \sim \frac{(m_g M/M_P^2)^{1/5}}{m_g}.
\eeq 
For a black hole 
of mass M, 
for instance, this means that we cannot trust 
perturbation theory in massive gravity within $d_v$, 
a distance much larger than the
black hole's Schwarzschild radius (here we implicitly assumed 
the experimental constraint that the graviton mass is extremely small).
This means physically that the nonlinear interactions 
associated with (massive) gravity are so large that, unlike the massless 
case, they can no longer be neglected at the scale $d_v$.
In the DGP model, as it was noted in \cite{Deffayet:2001uk}, the gravitational fluctuation 
around a heavy source contains a large term. 
Although this term is pure gauge at 
the linear level, it does contribute to the nonlinear corrections, which subsequently 
grow bigger than the linear fluctuations within a distance  
\beq
d_{DGP}\sim (M L^2/M_P^2)^{1/3}.
\eeq
In this paper, we investigate how the above feature is changed in the regularized model. 
First, we review the main features of the model in the formalism 
of~\cite{a9}, taking codimension N=1 for simplicity.
Then, we investigate how the model reacts to a heavy source located at the
brane. We will see that, in the regularized version, there is no blow-up 
mode in the linearized solution, nor in the lowest-order (cubic)
nonlinear interaction term, 
and that we can thus 
trust perturbation theory up until phenomenologically acceptable scales. 
The solutions \textit{do} have the famous vDVZ discontinuity \cite{vdvz}, 
which means that the
theory will give different predictions than standard Einstein gravity in the bending of 
light around the sun for instance, unless one introduces new dynamics to cure 
the discontinuity (using a repulsive vector force, for instance). 

Finally we go to the effective field theory description of all modes and look for modes 
which get strongly coupled at energies below the cutoff.
We will see that the induced action on the brane 
is highly non-local in terms of the metric due to the non-trivial boundary conditions implied 
by the action
(\ref{sd}). Unlike the DGP case, all interacting 
modes receive healthy kinetic terms, signaling that
the strong coupling problem effectively does not occur in our framework. 
So, our regularization has traded the strong coupling problem for the 
``almost locality" (in terms of the gravitational 
degrees of freedom) of DGP \footnote{DGP is local in the sense that at 
least the highest dimensional
kinetic term operator is local in the gravitational fields.}. 
This is closely related to the behavior of 
``soft gravity'' studied in ref~\cite{Gabadadze:2003ck}.
We will comment on this and other features in the conclusion.
  
     
\section{A Scalar Toy Model}
Consider a scalar field described by the following action:
\beq \label{1.1}
 S= -\epsilon \int_{z<\Delta} d^5 x \ [ \de_M \varphi \de^M \varphi ]-\int_{z>\Delta} d^5 x  
\ [ \de_M \varphi \de^M \varphi ].
\eeq
We are interested in the dynamics on the hypersurface $z=0$. For that purpose we can 
calculate 
the effective four dimensional action for the scalar field $\varphi_0= \varphi(0)$ by 
integrating 
out the bulk \cite{a9}. 
This procedure amounts to solving the bulk equations for $\varphi$ with the constraint 
$\varphi(0)=\varphi_0$, and using this solution in the induced Lagrangian.
Inside the ``dielectric" (i.e. the region $z<\Delta$),
the solution to the bulk equations of motion is (we work 
in the four-dimensional Fourier transform  for notational simplicity):
\beq \label{phi}
 \varphi^I_p(z)= \cosh(p z) \varphi^1_p+\sinh(p z) \varphi^2_p;
\eeq
Outside the dielectric the solution which vanishes at infinity is:
\beq
 \varphi^O_p(z)= \varphi^3_p e^{-p z}.
\eeq
We know that $\varphi^1_p$ is
the boundary (four dimensional) field, which we keep fixed in our formulation. 
Solving the boundary condition at $z=\Delta$, 
we get:
\beq
 \varphi^2_p=-\frac{1+ \epsilon \tanh(p \Delta)}{\epsilon +\tanh(p \Delta)} \varphi^1_p.
\eeq
Now the induced action at $z=0$ is the boundary term obtained by partial integration of
equation (\ref{1.1}):
\beq
 S_{brane} (\varphi) = \epsilon \int d^4 x \ [ \varphi_p(z) \de_z 
                        \varphi_p(z) ] \ar.
\eeq
Since $\de_z \varphi_p(z)=p \varphi^2_p$ and by going back to position space
we obtain the following induced action:
\beq
 S_{brane} (\varphi) = -\epsilon \int d^4 x \ \varphi_0 \sqrt{-\Box_4} \left[ 
 \frac{1+ \epsilon \tanh( \sqrt{-\Box_4} \Delta)}
{\epsilon +\tanh( \sqrt{-\Box_4}\Delta)} \right] \varphi_0.
\eeq
This action looks highly non-local, but we can use it as an effective field theory 
with cutoff $\Lambda_{UV}=\Delta^{-1}$.
Then we see that for $p \Delta\ll 1$ the action is:
\beq
 S_{brane} (\varphi) = L \int d^4 x  \ \varphi_0 \left( \Box_4 
-\sqrt{-\Box_4} /L \right) \varphi_0.
\eeq
So what is the net effect of the five dimensional dielectric on the four dimensional 
induced 
action? There appear two regimes on the brane, defined by the "crossover scale" 
$L= \epsilon \Delta$.
In the regime $p L \gg 1$, there is four dimensional propagation with propagators scaling
 as $1/p^2$, 
whereas in the complementary regime $p L \ll 1$, the propagation is five dimensional. So by 
this mechanism
we get the same propagation as in the DGP model. The crucial difference here is that we 
do not have to 
include a separate four dimensional kinetic term on the brane to get four dimensional
 behavior within
a certain range at the brane. It is induced by the bulk Lagrangian itself. 


\section{The Vainshtein Scale}
We look at the gravitational action on a half 
space, in which there are two bulk regions
to be distinguished: inside and outside the "dielectric". The gauge-fixed bulk action reads:
\beq \label{action}
S_{Bulk}/M_*^3= \epsilon S_{z<\Delta} +S_{z>\Delta},
\eeq
where $S_{z\lessgtr\Delta}$ is given by:
\beq\label{action'}
 S_{z\lessgtr\Delta}= \int_{z\lessgtr\Delta} d^5 x  
\sqrt{\gamma} \frac{1}{N} \left[R(\gamma)- 
K_{\mu \nu} K^{\mu \nu}+K^2 - \frac{1}{2} F_M^2\right].
\eeq
Here $K^{\mu \nu}$ is the extrinsic curvature and we used the notation $N_{\mu}=g_{\mu 5}$
and $N=g_{55}^{-1/2}$. The mass scale $M_*$ is the bulk Planck scale ($\sim 10^{-3}$eV-$100$MeV),
and $\epsilon= M_P^2/M_*^2 $. For $\Delta$ we choose the natural value $M_*^{-1}$. 
Finally, we choose the de Donder gauge fixing in the bulk
$ F^M \equiv \de_N (\sqrt{g} g^{MN})$. At the 
linearized level, expanding $g_{MN}= \eta_{MN}+h_{MN}$, this action gives the usual 
bulk equations for the fluctuations: 
\beq \label{DD}
 \Box_5 (h_{MN}-\frac{1}{2} \eta_{MN} h_5) =0,
\eeq
This equation holds inside and outside the dielectric, and the two solutions of this 
equation
should be matched over the surface using the appropriate boundary conditions. 
Varying the action in the two regions gives separate contributions to the interface
$z=\Delta$, which give us the following conditions [we use the notation $\lfloor J \rfloor = \epsilon 
J(\Delta_-)- J(\Delta_+)$]:
\beq \label{1.1a}
  \lf  K_{\mu \nu} - \eta_{\mu \nu} K -\h \eta_{\mu \nu} F_5 \rf=0,
\eeq
\beq \label{1.2}
  \lf F_{\mu} \rf =0,
\eeq
\beq \label{1.3}
  \lf F_5 \rf =0.
\eeq
These equations are obtained by variation of 
$h_{\mu \nu}$,
$N_\mu$ and $N$ respectively. $F_M$ appears here in its familiar linearized form; 
$F^N=\de_M (h^{MN}-\h \eta^{MN} h_5)$.
Using for instance the trace of equations (\ref{1.1a}) and (\ref{1.3}), we obtain ($\dot{x}=\de_z x $):
\beq 
 \lf \dot{h}_{55} \rf =0. 
\eeq
On the other hand, equation (\ref{1.1a}) gives:
\beq \label{dd}
 \lf \dot{h}_{\mu \nu}-2 \de_{(\mu} N_{\nu)} \rf =0.
\eeq
If we assume a source which is localized at the brane, the boundary conditions at 
the brane are given by the following set of equations:
\beq \label{2.1}\left.
 \left[\dot{h}_{\mu \nu}-2 \de_{(\mu} N_{\nu)}- \eta_{\mu \nu} \left( \h\dot{h}_4- \de.N+
 \h \dot{h}_{55} \right) \right] \right|_0= 
 -\frac{1}{2} \frac{T_{\mu \nu}}{\epsilon M_*^3},
\eeq
\beq \label{2.2}
 \left.\left[\de^{\kappa} h_{\mu \kappa}+ \dot{N}_{\mu}-\h \de_{\mu}\left( h_4
+ h_{55}\right)\right]\right|_0=0 ,
\eeq 
\beq \label{2.3}
 \left. \left[\de_{\mu} N^{\mu}+\h \dot{h}_{55}-\h \dot{h}_4 \right]
\right|_0=0 . 
\eeq
These are the initial conditions on the set of gravitational fields. 
Now we notice that the bulk linearized equations of motion for $N_\mu$ and $h_{55}$, 
$\Box_5 N_\mu =0$ and $\Box_5 (h_{55}-h_4)=0$ respectively, allow us to make the following
consistent gauge choice;
$N_{\mu}=0$ and $h_{55}=h_4$ (this is in specific consistent with the boundary conditions given by equations
(\ref{2.1})-(\ref{2.3})). This means we can essentially treat 
this as a scalar system at the linearized level, with in particular the fluctuation 
obeying:
\beq 
 \lf \dot{h}_{\mu \nu} \rf =0. 
\eeq
Assume now that we couple gravity to a source on the brane. What
would the solution for $h_{\mu \nu}$ look like then? 
The boundary condition at the brane is:
\beq
\left. \left[\dot{h}_{\mu \nu}-\eta_{\mu \nu} \dot{h}_4 
\right]\right|_0= -\frac{1}{2} \frac{1}{\epsilon M_*^3} 
T_{\mu \nu}.  
\eeq
This leads to a solution of the form (in four-dimensional momentum space):
\beq \label{lin}
 h_{\mu \nu}(p,z)= -\frac{1}{2} D(p,z) \left[ T_{\mu \nu}(p) -\frac{1}{3} 
  \eta_{\mu \nu} T(p) \right].
\eeq
Here $D(p,z)$ is the bulk to boundary scalar propagator of the theory, which is calculated
in the appendix. The point is made by contrasting the solution obtained in the DGP model, using the 
de Donder gauge in the bulk:
\beq\label{lin'}
 h_{\mu \nu}(p,z)=-\frac{1}{2} D'(p,z)
 \left[ T_{\mu \nu}(p) -\frac{1}{3} 
  \eta_{\mu \nu} T(p) -\frac{L}{3 p} p_{\mu} p_{\nu} T(p) \right],
\eeq 
where $D'(p,z)= -\frac{1}{M_P^2}\frac{1}{p^2+p/L} e^{-pz}$. 
The difference is clear: in the DGP case we get an extra term
 which blows up rapidly  with increasing $p$, while this term is absent in the regularized 
model. For DGP, this 
mode does not contribute at linear order, since it is pure gauge, but 
it does contribute at quadratic
order (cubic terms in the Einstein-Hilbert action). This 
means that in DGP, as in 
massive gravity, the linearized approximation breaks down at a 
distance scale much higher than the massless case. 
In DGP and massive gravity, the contribution 
from the gravitational radiation 
to the  energy density is much bigger than in the massless case, 
so that the heavy back-reaction in these models is not negligible. We 
discuss this in more detail now. 

A source confined to the brane, $ T_{MN}(x,z)= \delta(z) \delta_M^\mu 
\delta_N^\nu T_{\mu \nu} (x)$ 
gives rise at
linear level to $h^{(1)}(T)$, given by equation (\ref{lin}). This fluctuation 
contributes now by 
itself to the 
energy-momentum tensor $T^{(2)}_{MN}(x,z)=h^{(1)}(T) G h^{(1)}(T)$, with $G$ some operator 
with complex tensor 
structure but
quadratic in the derivatives. This gravitational radiation is now \textit{everywhere} 
in space, not just confined to the brane. Perturbatively we can now solve for 
$h^{(2)}$, which 
will be quadratic in the 
source. There are obviously two possibilities. The first one, like in standard 
Einstein gravity, 
is that $h^{(2)}$ is 
very small in comparison to  $h^{(1)}$, so that perturbation theory can be trusted,
 and one can 
keep the linearized
approximation up until, say,  the Schwarzschild radius of a black hole. The second 
possibility is 
that $h^{(2)}$ grows
bigger than $h^{(1)}$ at some large distance scale. This can be due \textit{either} to 
some 
large pieces in $h^{(1)}$ which are pure gauge only at linear order [and can be present 
in physical contributions to $h^{(2)}$], 
\textit{or} because the propagator of the system contains large pieces, which, applied to 
$T^{(2)}_{MN}$ generate fast growing terms. 

Instead of dwelling on this point, we may as well compute the second order 
fluctuations and see if either one
of the possibilities is present. 
To solve for this system in the de Donder gauge we need the scalar propagator, 
which is calculated in the
appendix. The equations of motion take the following form:
\beq
 \Box_5 (h^{(2)}_{MN}-\h \eta_{MN} h^{(2)}_5)= -\frac{1}{2}k_i T^{(2)}_{MN}(x,z),
\eeq
where the coupling $k_i$ is different on both sides of the interface. 
Obviously the extra constraints $N^{(2)}_{\mu}=0$ and $h^{(2)}_{55}=h^{(2)}_{4}$ are not
 automatically satisfied, but we can use 
residual gauge invariance to put:
\beq
 N^{(2)}_\mu\arrowvert_\Delta=0,
\eeq 
\beq
h^{(2)}_4\arrowvert_\Delta=h^{(2)}_{55} \arrowvert_\Delta.
\eeq 
Indeed, any coordinate transformation $\xi_{M}(x,z)$ obeying $\Box_5 \xi_{M}(x,z)=0$ will leave
the equations of motion invariant (but not the boundary conditions), 
and they can be used to enforce the above identities. 
Since $\delta h_{MN}=2\de_{(M} \xi_{N)}$ is a gauge transformation, 
it will not contribute to any amplitude between sources since we assume them conserved. We will discuss
these transformations in more detail below.  
In any case, we see that  the boundary conditions for $h^{(2)}_{\mu \nu}$ and $h^{(2)}_{55}$ are now the 
same as before, so that the scalar propagator still determines
the solution uniquely: 
\beq
h^{(2)}_{M N}=\frac{1}{2} \int d^5 x' D(x-x';z,z') \left[ T^{(2)}_{MN}(x', z')
-\frac{1}{3} \eta_{M N}T^{(2)}(x', z') \right].
\eeq
In other words, if neither the scalar propagator $D(x-x';z,z')$ (see appendix) nor the source
grow big anywhere, the second order solution does not grow big anywhere. 
So we can be sure that the linearized approximation does not break down at any 
unacceptably low scale. 

In the preceding derivation, we have assumed that the gauge transformations we use are globally
defined, i.e. that they are \textit{proper} gauge transformations. One might suspect that in fact 
the interface will bring complication, and one has to check that the gauge degrees of
freedom truly decouple \textit{globally} (from $T^{(2)}$). This amounts to checking if the following 
integral vanishes:
\beq
 S=\int_{z<\Delta} d^5 x \ \de^{(M} \xi^{N)} T^{(2)}_{M N}+ \int_{z>\Delta} d^5 x \ \de^{(M} 
\xi^{N)} T^{(2)}_{M N}. 
\eeq 
Since gravitational radiation is locally conserved:
\beq
 \de^{M} T^{(2)}_{M N}=0,
\eeq
the only contributions left are exactly the ones at the interface: 
\beq
S_{\mu}=  \int d^4 x \ \xi^{\mu} T^{(2)}_{\mu 5} \arrowvert_{\Delta-}-\int d^4 x \ \xi^{\mu} 
T^{(2)}_{\mu 5} \arrowvert_{\Delta+},
\eeq
\beq
 S_{5}= \int d^4 x \ \xi^5 T^{(2)}_{55} \arrowvert_{\Delta-}-\int d^4 x \ \xi^5 T^{(2)}_{55}
 \arrowvert_{\Delta+}.
\eeq
In the first integral, we can calculate the $T^{(2)}_{\mu 5}$ directly. Since $N^{(1)}_{\mu}=0$ 
everywhere, 
the expression for $T^{(2)}_{\mu 5}$ is particularly simple. The reader can 
check that, indeed:
\beq
 T^{(2)}_{\mu 5} \sim R^{(2)}_{\mu 5},
\eeq
where $R^{(2)}_{\mu 5}$ is the part of the Ricci tensor quadratic in $h^{(1)}_{\mu \nu}$. The 
constant of proportionality is the inverse coupling constant:
$\epsilon M^{3}_* $ inside the dielectric, and $M^{3}_* $ 
outside of it. Explicitly:
\begin{eqnarray}
 R^{(2)}_{\mu 5}=-\frac{1}{2} h^{\lambda \nu} \de_{\mu} \de_{5} h_{\lambda \nu}+\frac{1}{2} h^{\lambda \nu} 
 \de_5 \de_{\lambda} h_{\mu \nu} 
 +\frac{1}{2} \de_{\lambda} h_{55} \de_5 h^{\lambda}_{\mu} \nonumber \\
 -\frac{1}{4} \de_5 h_{M N} \de_{\mu} h^{MN}.
\end{eqnarray}
Now we notice that all terms appearing are linear in $\de_5 h_{MN}$, which 
means:
\beq
  T^{(2)}_{\mu 5}\arrowvert_{\Delta-}=\epsilon M^{3}_*  R^{(2)}_{\mu 5}\arrowvert_{\Delta-}
  =M^{3}_*  R^{(2)}_{\mu 5}\arrowvert_{\Delta+}=T^{(2)}_{\mu 5}\arrowvert_{\Delta+}.
\eeq
This was expected if our system was to be stable, of course: 
any inequality here would imply
nonzero pressure on either side of the dielectric. Since we use continuous gauge transformations $\xi_{\mu}$,
 we can conclude that $S_{\mu}=0$.

As for $S_5$, we only have to notice that $\xi_5$ vanishes at the boundary. Indeed, since we
demanded that $h_{55}$ is continuous over the boundary and 
$\lf \dot{h}_{55} \rf =0$, we can conclude
that the gauge parameter $\xi_5$ (which transforms $\delta h_{55} = 2 \de_5 \xi_5$) obeys $\lf \xi_5 \rf=0$, 
implying $\xi_5 \arrowvert_{\Delta} =0$.

The vanishing of $S_5$ and $S_{\mu}$ also ensures that, at cubic order in 
the interactions, all the ghost degrees of freedom
decouple. It would be interesting to see if this property persists to all order. 

\section{Mode Analysis}
We now turn to the mode analysis of the complete four-dimensional induced 
Lagrangian. Before we start, it is a good idea to review shortly what happens
in the case of massive gravity and DGP. 

In massive gravity, the only local, ghost-free, free 
Lagrangian is given by adding the Pauli-Fierz
term to the linearized Einstein-Hilbert action:
\beq
 \mathcal{L}_{PF}= M_P^2 \sqrt{g} R^L(g)\arrowvert_{g=\eta+h} +m_g^2 M_P^2(h_{\mu \nu}^2 -h^2).  
\eeq
The Pauli-Fierz term breaks coordinate invariance explicitly, and this Lagrangian
propagates five degrees of freedom. One can restore this invariance by 
introducing a
St\"uckelberg field, by the substitution 
$h_{\mu \nu} \to h_{\mu \nu}+2\de_{(\mu} A_{\nu)}$, and 
demanding that under coordinate transformations $\delta A_{\mu}=-\xi_\mu$ (\cite{FCV},\cite{Schwartz}). 
We decompose $A_{\mu}$ in a longitudinal and transverse part $A_{\mu}=A^T_{\mu}+\de_{\mu}a$. 
$A_{\mu}$ is essentially the Goldstone field of broken coordinate invariance, whose
proper nonlinear definition was given in \cite{a0}. 
This substitution $h_{\mu \nu}$ leaves the Einstein-Hilbert action invariant, as 
it is invariant under coordinate transformations, but in the mass term $A_{\mu}$ appears in 
the following form:
\beq
 \mathcal{L}_{mass}=m_g^2 M_P^2 [F_{\mu \nu}^2+2A_{\mu, \nu} h^{\mu \nu}-2 
A_{,\mu}^\mu h_\nu^\nu], 
\eeq 
where $F_{\mu \nu}$ is the field strength of $A_{\mu}$. One now immediately sees that the 
transverse part of the field has the correct kinetic term, but that the longitudinal scalar
gets a kinetic term only by mixing with $h_{\mu \nu}$. The resulting kinetic term, after 
diagonalization, reads:
\beq
 \mathcal{L}_a= m_g^4 M_P^2 a \Box_4 a.
\eeq  
Now the logic is as follows. The field $a$ is expected to obtain its kinetic term only 
by mixing
because otherwise it would obtain a higher derivative kinetic term, implying violation 
of unitarity.
The kinetic term it obtains is anomalously small as a consequence of the mixing, 
which means that
the higher dimensional operators containing $a$, which are generated in the interacting 
theory, could
actually become important at a very low scale. To see when these interactions become 
important we have 
to use the canonically normalized field $a_c=a/(m_g^2 M_P)$. 
At the quantum level, the only symmetry protecting operators being created 
is the shift symmetry mentioned above. Thus we expect for 
instance the appearance of interactions of the form:
\beq
 \mathcal{L}_{int} \supset  \frac{1}{m_g^4 M_P} (\de^2 a_c)^3. 
\eeq 
This term can also be seen to appear explicitly in the self-interaction 
terms of the Goldstone boson~\cite{a0}.
This interaction is in fact the strongest interaction that will appear in the Lagrangian, 
 and it becomes strong at $\Lambda_5= (m_g^4 M_P)^{1/5}$. At this scale we
 should include all
operators consistent with the symmetries, and we need a UV completion to regain
predictability above that scale. 

For DGP, the situation looks vastly different at first. The theory obeys manifestly 
general covariance, and  
hence it does not share the arbitrariness of massive gravity. Recall that 
the Pauli-Fierz term  is only uniquely defined at   
quadratic order in the fluctuations, and its nonlinear extensions are quite 
arbitrary. 
Specifically, the generation of quantum interactions beyond quadratic order 
is not protected by general covariance.   
The induced gauge-fixed action at the brane takes the form~\cite{a9}:
\begin{eqnarray} \label{dgp}
 \mathcal{L}_{DGP}=M_P^2 \Big[
\frac{1}{2}h^{\mu \nu}(\Box_4-m\triangle)h_{\mu \nu}
        -\frac{1}{2} h_4(\Box_4-m\triangle)h_4    &    \nonumber \\ 
         - m N^{\mu} (\triangle+m) N_{\mu}
        +\frac{m}{2} h_4 \triangle h_{55}-
         m N^{\mu} \de_{\mu} h_{55}
        -\frac{m}{4} h_{55}\triangle h_{55} \Big],
\end{eqnarray}
where we used the notation $\triangle=\sqrt{-\Box_4}$, 
and where $m=1/L=M_*^3/M_P^2$.
By analyzing the scalar modes in this Lagrangian , 
one sees that there is indeed a scalar mode $\pi$,
defined by $N_{\mu}=\de_{\mu} \pi, h_{55}=-2 \triangle \pi$ \cite{a9}, which  gets a kinetic term
only by mixing 
with the trace part of $h_{\mu \nu}$. The mixing term in the action (\ref{dgp}) is proportional to:
\beq
  M_P^2 m \pi \Box_4 h_4. 
\eeq
This means that after diagonalization this mode will receive a kinetic term:
\beq
 \mathcal{L}_{\pi} \sim M_P^2 m^2 \pi \Box_4 \pi. 
\eeq
Again we see that the kinetic term is suppressed. 
Now, the bulk interactions induce the following cubic interaction at the brane:
\beq
\mathcal{L}_{int}= M_P^2 m \de^{\mu} \pi \de_{\mu} \pi \Box_4 \pi. 
\eeq
Using the canonically normalized field, $\pi_c = \pi/(M_P m)$, one can easily
compute the scale at which this interaction 
becomes important: $\Lambda_3= (M_P/L^2)^{1/3}$. 
Unlike in massive gravity, this operator has canonical dimension seven, it is
explicitly present in the classical Lagrangian and is not renormalized. 
The mode has, by its definition, the 
physical interpretation of being the ``brane bending.''  

We are now in the position to investigate the appearance of a strong coupling problem 
by following the procedure used in  the 
previous examples. We will calculate the induced Lagrangian of our system and 
investigate all its modes. 
We may expect some scalars to receive their kinetic terms by mixing, 
especially the brane bending mode. If this mixing
is indeed small, we can conclude that our model shares the same problem as massive 
gravity and DGP. 

In order to calculate the induced Lagrangian, we need the on-shell bulk 
solutions of each mode (in the de Donder gauge), with all the fields 
constrained by the boundary conditions given by equations (\ref{1.1a}-\ref{1.3}). 
Using the orthogonal decomposition of the four dimensional fluctuation into 
its transverse-traceless, longitudinal, and trace parts~\footnote{This 
decomposition is unique because 4-D fields that are transverse and 
longitudinal at the same time have vanishing 4-D momentum. So, they do not
extend to 5-D on-shell fields vanishing at $z=\infty$.}: 
\beq
 h_{\mu \nu}= h^{TT}_{\mu \nu} + 2 \de_{(\mu} \zeta_{\nu)} +\eta_{\mu \nu} \phi, 
\eeq
we get the following (mixed) boundary condition  from equation (\ref{1.1a}): 
\beq
 \lf \dot{h}^{TT}_{\mu \nu}+\eta_{\mu \nu} \dot{\phi}- 2 \de_{(\mu} 
\eta_{\nu)} \rf =0,
\eeq
where we defined for later convenience $ \eta_{\mu}\equiv N_{\mu}-\dot{\zeta}_{\mu}$.
Taking the trace of this and comparing with the double (four dimensional) divergence 
of equation (\ref{dd}) we see that:
\beq
 \lf \dot{\phi} \rf =0.
\eeq
Now it is easy to obtain the remaining boundary conditions on the fields. For the 
transverse traceless  part we have:
\beq
 \lf \dot{h}^{TT}_{\mu \nu} \rf =0. 
\eeq
For $N_{\mu}$ and $A_{\mu}$ we essentially get conditions on their linear combination
 $\eta_{\mu}$:
\beq \label{et}
 \lf \eta_{\mu} \rf =0,
\eeq
\beq \label{ctu}
 \lf \dot{\eta}_{\mu} -\de_{\mu} \phi \rf = 
\left. \frac{\epsilon}{2} \de_\mu h_{55} \right|_\Delta . 
\eeq
Here, we kept only the leading term in an $\epsilon\gg1$ expansion. 
So, we notice that the transverse and trace part of 
$h_{\mu \nu}$ as well as $h_{55}$ 
obey homogeneous boundary conditions, while $N_{\mu}=N^T_{\mu}+\de_{\mu} \theta$ and $A_{\mu}$
 obey mixed boundary conditions. 
Furthermore, we ask continuity of the fields $h_{\mu \nu}$, $h_{55}$ and $\dot{N}_{\mu}$ 
(this last constraint follows
from imposing the gauge fixing). 
For a field obeying homogeneous boundary conditions (continuity of $J$ and $ \lf \dot{J} \rf =0$),
 the solution 
is the same as the scalar example discussed in sections 2 and 3.   
Those fields thus include the transverse
traceless part of $h_{\mu \nu}$, $h^{TT}_{\mu \nu}$, and two scalars, $\phi$ and
$h_{55}$. 
We proceed by solving for $\eta$. It obeys equation (\ref{et}) and has a continuous 
derivative over the interface. 
Setting $\eta_{\mu}=\eta^T_{\mu}+\de_{\mu} \eta$, we see that only the longitudinal part
obeys non-homogeneous boundary conditions, and we can solve it by:
\begin{eqnarray}
 \eta= \eta_I \sinh(p z+b) && z<\Delta, \\
 \eta= \eta_o e^{-p z}     && z>\Delta.
\end{eqnarray}
We easily find that $b=-1/\epsilon - p \Delta$. By equation (\ref{ctu}) 
and the continuity of $\phi$, $h_{55}$ and $\dot{\eta}$, we see that these 
fields are related by:
\beq \label{approx}
h_{55} \arrowvert_0 \approx h_{55} \arrowvert_{\Delta} 
\approx 2(\dot{\eta}-\phi)
\arrowvert_{\Delta} \approx 2(\dot{\eta}-\phi)\arrowvert_{0}.
\eeq
This constraint already signals a major difference between our regularization
and DGP in the de Donder gauge. In DGP, the boundary theory has three 
independent scalar fields, while here there are only two [see eq.(~\ref{dgp})
and ref.~\cite{a9}].
 
The induced brane action obtained by integrating out the bulk linearized 
action (\ref{action}) reads:
\begin{eqnarray}
 \Gamma_{ind} & = & -\frac{\epsilon M_{*}^3}{4} \int d^4 x \Big[ 
-h_{\mu \nu} \dot{h}_{\mu \nu}
  -\frac{1}{2} h_{55} \dot{h}_{55}+\frac{1}{2} h_{4} \dot{h}_{4}-2 N_{\mu} \dot{N}^{\mu} \nonumber \\
 & & +\frac{1}{2} (h_{55} \dot{h}_4+ h_{4} \dot{h}_{55})-2 N_{\mu} (2 \de_{\kappa} h^{\kappa \mu} - \de^{\mu} h_4 
 -\de^{\mu} h_{55})\left. \Big]\right|_0
\end{eqnarray}
The transverse traceless part and the transverse vector in the metric decouple from the other modes.  
Concentrating on $h^{TT}_{\mu \nu}$ for instance, it will appear in the action only with the term:
\beq\label{tensact}
 \Gamma_{ind} (h^{TT}) = \frac{M_P^2}{4} \int d^4 x \ \left[ h^{TT \mu \nu} 
                     (\Box_4-\triangle/L )
                       h^{TT}_{\mu \nu} \right].
\eeq
We further concentrate on the scalar sector of the theory. 
Substituting all the projections of the field, one obtains after some algebraic manipulations:
\begin{eqnarray}
\Gamma_{ind} (\zeta,\phi,h_{55},\theta)=-\frac{\epsilon M_{*}^3}{4} \int d^4 x \ [-2 \zeta \Box^2 \dot{\zeta}
+2 \phi \Box \dot{\zeta} 
  +2 \dot{\phi} \Box \zeta + 4 \phi \dot{\phi} \nonumber \\
- \frac{1}{2} h_{55} \dot{h}_{55} +h_{55} \Box \dot{\zeta} 
+\dot{h}_{55} \Box \zeta 
+2 h_{55} \dot{\phi}+2 \dot{h}_{55} \phi+ \nonumber \\
2 \theta \Box \theta +4 \theta \Box^2 \zeta 
 -4\theta \Box \phi -2 \theta \Box h_{55}]. \label{scalact}
\end{eqnarray}
We now define:
\beq
 \psi= \dot{\eta} = \dot{\theta}+\Box_4 \zeta,
\eeq 
where the last equation follows from the 5-D on-shell condition. 
We can eliminate $h_{55}$ from the action using the constraint (\ref{approx}), so that the action gets the 
simple form:
\beq\label{simpact}
 \Gamma_{ind} (\psi,\phi)=-\frac{\epsilon M_{*}^3}{4} \int d^4 x \ [-6 \phi \dot{\phi}+6 \dot{\psi} \phi
 +6 \psi \dot{\phi}].
\eeq
We notice that the field $\psi$ does not appear to have a diagonal 
kinetic term. Since the fields $\psi$ and $\phi$
 have the same behavior [in momentum space $\sim A \cosh(-1/\epsilon-p\Delta+pz)$], the last two
terms are identical, and we are ready to write the final form of the action using the solution of both fields 
in the bulk;
\beq
 \Gamma_{ind} (\psi,\phi)= \h \int d^4 x \ [ 3 M_P^2 \phi(\Box_4- m \triangle) \phi+6\epsilon M_{*}^3 \eta \Box_4 \phi].
\eeq
Indeed, the scalar $\eta$ is the one we set out to find from the beginning. As in both DGP and massive gravity, it gets 
its kinetic term by mixing with the scalar part $\phi$ of the metric. But unlike in DGP or massive gravity, the mixing 
term is \textit{very large}! It will not give rise to an anomalously small kinetic term, but a very big kinetic term for the 
field $\eta$;
\beq
 \mathcal{L}_{\eta} \sim  M_P^2 M_{*}^2 \eta \Box_4 \eta.
\eeq 
The canonically normalized field $\eta_c= \eta/(M_P M_*)$ will not cause trilinear 
interactions in the theory to blow up at unacceptably low energy scales. As a 
rough estimate, we can see that the most dangerous induced interaction will give a cubic
coupling of the form:
\beq
\mathcal{L}_{cub} \sim  \frac{1}{M_P M_*^2} \ \eta_c p^2 \eta_c p^2 \eta_c,
\eeq  
which gives a strong coupling scale:
\beq
 \Lambda_3=(M_P M_*^2)^{1/3} \gg M_*.
\eeq
So we can trust our framework up until the cutoff we imposed, $M_*$.

\section{Conclusion}
The results obtained in this paper are puzzling. We could have thought that our
regularization of the DGP model differs from the unregulated version only at
distances $D\ll \Delta$. Instead, we have seen that it differs even at the
vastly larger scale $\Lambda_{DGP}^{-1}$. How is this possible, and, where 
are potential problems hidden?

The origin of this difference can be seen in two ways. First, by comparing
eq.~(\ref{lin}) with the DGP result eq.~(\ref{lin'}). In eq.~(\ref{lin}),
unlike in the DGP case, the trace of the metric fluctuation, $h_\mu^\mu$, 
propagates and is a ghost. This is not a disaster because the difference 
between the DGP propagator and that of the regularized theory is a gauge 
transformation. The absence of a gauge mode, that becomes large at the scale 
$d_v$, is also a feature of the ``softly massive'' gravity studied in 
ref.~\cite{Gabadadze:2003ck}.
So, as long as linear gauge transformations decouple, all 
ghosts decouple. In section 3, we explicitly checked that this indeed happens
to cubic order in the interactions. Clearly, an all-order check of this
property is crucial for the ultimate viability of our regularization.

The other way of understanding the difference between DGP and the model 
presented here is to notice that the linearized action 
eqs.~(\ref{tensact},\ref{scalact}) is {\em very} nonlocal when expressed 
in terms of $h_{\mu\nu}$.
The free DGP action is local in the UV, since its highest-dimension term
is local. The action in eqs.~(\ref{tensact},\ref{scalact}), instead, is never
local because of the implicit presence of spin-projection operators.
By this we mean the following: a high-spin theory is always nonlocal, when 
written in terms of just its physical degrees of freedom. In massive gravity
or DGP, one can add unphysical polarizations --a massless vector and 
a scalar for each helicity $\pm 2$ polarization-- so that the 
free kinetic term is local. In our case, this is not possible, because of the
uniqueness of the Pauli-Fierz action.

When written in terms of the spin-projected components, the action still
exhibits a potential problem due to the presence of ghosts. It is sufficient
to notice that in the free action eq.~(\ref{simpact}), the field
$\chi\equiv \dot{\theta}-\Box_4 \zeta$ never appears. It has zero kinetic 
term, so, if it appears at all in the nonlinear terms, it would interact with
infinite strength. Since $\chi$ is precisely a 4-D gauge transformation, 
our calculation in section 3 proves again that this does not happen at cubic 
order in the interactions. This gauge transformation only acts on the 4-D 
coordinates, while leaving the interface at $z=\Delta$ fixed. So, it is 
reasonable to think that it is non-anomalous, and that it will decouple at all 
orders. Again, an explicit proof of this property is important for the 
consistency of our regularization.
We hope to address this question in the near future.     
\subsection*{Acknowledgments}
M.P. is supported in part by NSF grants PHY-0245069 and PHY-0070787;
J.-W.R. is supported by an NYU Henry~M.~McCracken Fellowship.


\section*{Appendix A: The Scalar Propagator}
In this section, we calculate the scalar propagator for our system. We basically have 
to solve the 
following equation:
\beq \label{preq}
 (p^2-\de_z^2) \tilde{\phi} = \frac{1}{k_i} \delta(z-z').
\eeq
There will be two distinct cases: $z'<\Delta$, $k_i=\epsilon M_*^3$, and 
$z'>\Delta$, $k_i= M_*^3$. We treat them separately here. 

\subsection*{Case I: $z'<\Delta$}
 
There are three different regions; region I, which is for values $0<z<z'$ (we demand that 
$\dot{\phi} 
\arrowvert_0 = 0$), 
region II, which is for values $z'<z<\Delta$, and then region III, which is the space outside 
the dielectric.
The solutions to equation (\ref{preq}) are:
\begin{eqnarray}
 \phi_I(p,z)= A \cosh(p z), \nonumber\\
 \phi_{II}(p,z)= B \cosh(p z)+ C \sinh(p z), \\
 \phi_{III}(p,z)=D e^{-p (z-\Delta)} \nonumber.
\end{eqnarray}  
The continuity equations are:
\begin{eqnarray}
 A \cosh(p z') = B \cosh(p z')+ C \sinh(p z'), \\
 D= B \cosh(p \Delta)+ C \sinh(p\Delta) \nonumber,
\end{eqnarray} 
while the derivative boundary conditions read:
\begin{eqnarray}
p B \sinh( p z')+ p C \cosh(p z') - p A \sinh(p z')= \frac{1}{ \epsilon M_*^3},  \\
\epsilon \left[ p B \sinh(p \Delta) + p C \cosh(p \Delta) \right]= -D p \nonumber.
\end{eqnarray} 
Adding the continuity and derivative conditions we obtain, in the regime where
 $p \Delta \ll 1$ and
in the approximation $\epsilon\gg1$:
\begin{eqnarray}
C= -\left( \frac{1}{\epsilon} + \Delta p \right) B, \nonumber \\
A \cosh(p z') = B \left[ \cosh(p y') - 
\left( \frac{1}{\epsilon} + \Delta p \right) \sinh(p z')
 \right], \\
 B \left[ p \sinh(p z') - p^2 \Delta \cosh(p z') \right] -A p \sinh(p z') =
\frac{1}{\epsilon M_*^3} \nonumber.
\end{eqnarray}
The last two equations give us $A \approx B$ and:
\begin{eqnarray}
 A= -\frac{1}{ \epsilon \Delta M_*^3} \frac{1}{p^2 +\frac{p}{ \epsilon \Delta}}, \\
C= \frac{1}{\epsilon M_*^3 p}\nonumber.
\end{eqnarray}
This leads to the following propagator for $z'<\Delta$:
\begin{eqnarray}
 \phi_I(p,z)= \phi_{II}(p,z)=-\frac{1}{M_P^2} \frac{1}{p^2+p/L} \cosh(p z),\\
 \phi_{III}(p,z)=-\frac{1}{M_P^2} \frac{1}{ p^2 +p/L} e^{-p (z-\Delta)} \nonumber.
\end{eqnarray}  

\subsection*{Case II: $z'>\Delta$}
There are three different regions; region I, which is for values $0<z<\Delta$ (we demand 
that $\dot{\phi} \arrowvert_0 = 0$), 
region II, which is for values $\Delta<z<z'$, and region III, for which $z'<z$. 
Now the solutions are as follows: 
\begin{eqnarray}
 \phi_I(p,z)= A \cosh(p z),\nonumber\\
 \phi_{II}(p,z)= B \cosh(p z)+ C \sinh(p z), \\
 \phi_{III}(p,z)=D e^{-p (z-z')}\nonumber.
\end{eqnarray}  
The continuity and derivative equations read:
\begin{eqnarray}
 A \cosh(p \Delta) = B \cosh(p \Delta)+ C \sinh(p \Delta), \\
 D= B \cosh(p z')+ C \sinh(p z')\nonumber;
\end{eqnarray} 
\begin{eqnarray}
\epsilon p A \sinh(p \Delta) = p \sinh( p \Delta) B + p \cosh(p \Delta) C, \\
-D p - p B \sinh(p z') - p C \cosh(p z') = \frac{1}{M_*^3}\nonumber.
\end{eqnarray} 
Adding the derivative and continuity equations leads after some simplifications to 
$A \approx B$ and:
\begin{eqnarray}
 C \approx \epsilon \Delta p B, \\
 B= -\frac{1}{M_*^3} \frac{1}{p+p^2 L} e^{-p z'}\nonumber.
\end{eqnarray}
We write down the explicit solution for the propagator at $z'>\Delta$:
 \begin{eqnarray}
 \phi_I(p,z)= -\frac{1}{M_*^3} \frac{1}{p+p^2 L} e^{-p z'} \cosh(p z), \nonumber\\
 \phi_{II}(p,z)= -\frac{1}{M_*^3} \frac{1}{p+p^2 L} e^{-p z'} \cosh(p z)-\frac{1}{M_*^3} 
   \frac{pL}{p+p^2 L} e^{-p z'} \sinh(p z), \\
 \phi_{III}(p,z)= -\frac{1}{M_*^3} \frac{1}{p+p^2 L} \left[ \cosh(p z') + 
p L \sinh(p z') \right] e^{-p z}\nonumber.
\end{eqnarray}  
We see that in the case $ p z'\gg1$:
\beq
\phi_{III}(p,z)= -\frac{1}{M_*^3 p} e^{-p (z-z')}.
\eeq
The bulk to boundary propagator is obtained by putting $z=0$:
\begin{eqnarray}
\phi_I= \phi_{II}= -\frac{1}{M_P^2} \frac{1}{p^2+ p/L}, \\
\phi_{III}=  -\frac{1}{M_P^2} \frac{1}{p^2+ p/L}  e^{- p z'} \nonumber.
\end{eqnarray}
The propagator is well-behaved everywhere and does not grow anomalously large in any region. 



\begin{thebibliography}{99}


\bibitem{first}
R.~Gregory, V.~A.~Rubakov and S.~M.~Sibiryakov,
Phys.\ Rev.\ Lett.\  {\bf 84}, 5928 (2000)
[arXiv:hep-th/0002072].

\bibitem{dgp}
G.R.~Dvali, G.~Gabadadze and M.~Porrati, 
Phys.\ Lett.\ B {\bf 485}, 208 (2000) 
[arXiv:hep-th/0005016]; 
A.~Lue,
Phys.\ Rev.\ D {\bf 66}, 043509 (2002)
[arXiv:hep-th/0111168];
C.~Deffayet,
Phys.\ Lett.\ B {\bf 502}, 199 (2001)
[arXiv:hep-th/0010186].

\bibitem{ghost}
N.~Arkani-Hamed, H.~C.~Cheng, M.~A.~Luty and S.~Mukohyama,
arXiv:hep-th/0312099.

\bibitem{Dvali:2000xg}
G.R.~Dvali and G.~Gabadadze,
Phys.\ Rev.\ D {\bf 63}, 065007 (2001)
[arXiv:hep-th/0008054].

\bibitem{dgs}
G.~Dvali, G.~Gabadadze and M.~Shifman,
Phys.\ Rev.\ D {\bf 67}, 044020 (2003)
[arXiv:hep-th/0202174].

\bibitem{Gabadadze:2003ck}
G.~Gabadadze and M.~Shifman,
arXiv:hep-th/0312289.

\bibitem{Dubovsky}
S.L.~Dubovsky and V.A.~Rubakov,
Phys.\ Rev.\ D {\bf 67}, 104014 (2003)
[arXiv:hep-th/0212222].

\bibitem{KPR} 
M.~Kolanovic, M.~Porrati and J.-W.~Rombouts, 
Phys.\ Rev.\ D {\bf 68}, 064018 (2003) 
[arXiv:hep-th/0304148]. 


\bibitem{Kolanovic:2003da}
M.~Kolanovic,
Phys.\ Rev.\ D {\bf 67}, 106002 (2003)
[arXiv:hep-th/0301116].

\bibitem{dvali}
G.~Dvali and A.~Vilenkin,
Phys.\ Rev.\ D {\bf 67}, 046002 (2003)
[arXiv:hep-th/0209217].

\bibitem{a9}  
M.A.~Luty, M.~Porrati and R.~Rattazzi, 
JHEP {\bf 0309}, 029 (2003) 
[arXiv:hep-th/0303116]. 

\bibitem{Rubakov} 
V.A.~Rubakov, 
arXiv:hep-th/0303125. 

 
\bibitem{a0}   
N.~Arkani-Hamed, H.~Georgi and M.D.~Schwartz, 
Annals Phys.\  {\bf 305}, 96 (2003) 
[arXiv:hep-th/0210184]. 
 
\bibitem{vain}  
A.I.~Vainshtein, 
Phys.\ Lett. \ B {\bf 39}, 393 (1972). 
 
\bibitem{Deffayet:2001uk} 
C.~Deffayet, G.R.~Dvali, G.~Gabadadze and A.I.~Vainshtein, 
Phys.\ Rev.\ D {\bf 65}, 044026 (2002) 
[arXiv:hep-th/0106001]. 

\bibitem{vdvz} H van Dam and M. Veltman, Nucl. Phys. B {\bf 22} (1970) 397; 
V.I. 
Zakharov, JETP Lett. {\bf 12} (1970) 312; D.G. Boulware and S. Deser, 
Phys. Rev. D {\bf 6} (1972) 3368. 
 
\bibitem{FCV} 
M.~Porrati, 
Phys.\ Lett.\ B {\bf 534}, 209 (2002) 
[arXiv:hep-th/0203014]. 

\bibitem{Schwartz}
M.D.~Schwartz,
Phys.\ Rev.\ D {\bf 68}, 024029 (2003)
[arXiv:hep-th/0303114].

\end{thebibliography}
\end{document}